
\input phyzzx

\def\df{\varphi}

\def\bg{{\hat g}}
\def\bgr{{\rm e}^{2\rho} {\hat g}}
\def\bgs{{\rm e}^{2\sigma} {\hat g}}

\def\bcv{{\hat R}}
\def\vr{\delta \rho}

\def\vg{\delta g}
\def\vp{\delta \df}
\def\vf{\delta f}

\def\dg{\hbox{$\sqrt{-g}$}}
\def\dbg{\hbox{$\sqrt{-{\hat g}}$}}
\def\e{{\rm e}}

\def\pp{\prime}
\def\pd{\partial}

\REF\hwa{S. Hawking, Comm. Math. Phys. {\bf 43} (1975) 199; Phys. Rev. {\bf
D14}         (1976) 2460.}
\REF\dfu{P. Davies, S. Fulling and W. Unruh, Phys. Rev. {\bf D13} (1976) 2720.}
\REF\cghs{C. Callan, S. Giddings, J. Harvey and A. Strominger, Phys. Rev.
          {\bf D45} (1992) R1005.}
\REF\rst{J. Russo, L. Susskind and L. Thorlacius, Phys. Lett. {\bf B292}
         (1992) 13;
         T. Banks, A. Dabholkar, M. Douglas and M. O'Loughlin, Phys. Rev.
         {\bf D45} (1992) 3607.}
\REF\w{S. Hawking, Phys. Rev. Lett. {\bf 69} (1992) 406;
       L. Susskind and L. Thorlacius, Nucl. Phys. {\bf B382} (1992) 123;
       B. Birnir, S. Giddings, J. Harvey and A. Strominger, Phys. Rev.
       {\bf D46} (1992) 638.}
\REF\h{K. Hamada, Phys. Lett. {\bf B300} (1993) 322.}
\REF\ht{K. Hamada and A. Tsuchiya, {\it Quantum Gravity and Black Hole
        Dynamics in 1+1 Dimensions}, preprint UT-Komaba 92-14, to appear in
        Int. J. Mod. Phys. {\bf A}.}
\REF\his{W. Hiscock, Phys. Rev. {\bf D23} (1981) 2813.}
\REF\dk{J. Distler and H. Kawai, Nucl. Phys. {\bf B321} (1989) 509;
        F. David, Mod. Phys. Lett. {\bf A3} (1988) 1651.}
\REF\s{N. Seiberg, Prog. Theor. Phys. Suppl. {\bf 102} (1990) 319.}
\REF\tih{P. Thomi, B. Isaak and P. Hajicek, Phys. Rev. {\bf D30} (1984) 1168;
         P. Hajicek, Phys. Rev. {\bf D30} (1984) 1178.}

\nopubblock

\titlepage

\title{{\bf Quantum Gravity and Black Hole}\footnote\dagger{Talk
given by K.H. at \lq\lq Workshop on General Relativity and Gravity", Waseda,
Tokyo, Japan, 18-20 Jan 1993.}}

\author{Ken-ji Hamada and Asato Tsuchiya}

\address{Institute of Physics, University of Tokyo \break
         Komaba, Meguro-ku, Tokyo 153, Japan \break}

\centerline{\bf Abstract}

    The quantum theory of the spherically symmetric gravity in 3+1 dimensions
is investigated. The functional measures are explicitly evaluated and the
physical state conditions  are derived by using the technique developed in two
dimensional quantum gravity. Then the new features which are not seen in ADM
formalism come out. If $\kappa_s > 0 $, where $\kappa_s =(N-27)/12\pi $ and $N$
is the number of matter fields, a singularity appears, while for $\kappa_s <0$
the singularity disappears. The quantum dynamics of black hole seems to be
changed by the sign of $\kappa_s $.

{\bf 1. Introduction}

   Since the original work of Hawking,\refmark{\hwa} many authors study the
quantum dynamics of black holes. Almost all of works are done within the
semi-classical approximation.\refmark{\dfu,\cghs,\rst,\w}
In this talk we discuss how the quantum gravity will influence the dynamics of
black holes.\refmark{\h,\ht} As a model of gravity we consider the spherically
symmetric gravity in 3+1 dimensions.

  As a quantization method of gravitation, Arnowitt-Deser-Misner (ADM)
formalism is well-known. This method, however, has some serious problems,
which are the issues of measures and orderings. Here we explicitly
evaluate the contributions from measures. Following the procedure developed in
two dimensional quantum gravity we determine the measures in conformal gauge.
{}From the
gauge fixed theory the physical state conditions are derived. Then the new
features which are not seen in ADM formalism appear.

  The spherically symmetric gravity in 3+1 dimensions is defined by reducing
the Einstein-Hilbert action to two dimensional one as
$$
  I_{EH} = {1 \over 16\pi G} \int d^4 x \hbox{$\sqrt{-g^{(4)}}$} R^{(4)}
         = {1 \over 4} \int d^2 x \hbox{$\sqrt{-g}$}
                \biggl(R_g \df^2 +2 g^{\alpha\beta}
                         \pd_{\alpha} \df \pd_{\beta} \df +{2 \over G}
                   \biggr) ~.    \eqno(1)
$$
The fields $g_{\alpha\beta}$ and $\df $ are defined through the four
dimensional metric $ ( ds^{(4)} )^2 = g_{\alpha\beta}dx^{\alpha} dx^{\beta}
+ G \df^2 d\Omega^2 $, where $\alpha, \beta =0, 1 $ and $d\Omega^2 $ is the
volume element of a unit 2-sphere.  $G $ is the gravitational constant. In the
following we set $G=1 $.
We couple $N$ two dimensinal conformal matter fields
$$
     I_M (g,f) = -{1 \over 2} \sum^N_{j=1} \int d^2 x \dg
           g^{\alpha \beta} \partial_{\alpha} f_j \partial_{\beta} f_j ~.
      \eqno(2)
$$

   Some classical solutions of this system are known. For $f=0 $ the
Schwarzshild geometry is well-known. The gravitatinal collapse geometry is
given by\refmark{\his}
$$
       ds^2 = - \biggl( 1- {2M \vartheta({\bar v}) \over r} \biggr)
                {{\bar u} \over {\bar u}+4M \vartheta({\bar v}) }
                 d{\bar u} d{\bar v}  ~,
           \qquad  \df = r ~,
          \eqno(3)
$$
where $ds^2 =g_{\alpha\beta}dx^{\alpha}dx^{\beta} $ and the coordinate $({\bar
u},{\bar v})$ is
defined through the relations,
$d{\bar u} = du^{\star} ({\bar u}+4M)/{\bar u} $,
$u^{\star}=v-2r^{\star}$, $r^{\star}=r+2M \log ({r \over 2M}-1)$
and ${\bar v}=v $. This geometry is derived by sewing the flat space time and
the Schwarzshild black hole geometry along the shock wave line ${\bar v} =0 $,
where the infalling matter flux is given by
$T^f_{{\bar v}{\bar v}} = M\delta ({\bar v})$. In $({\bar u},{\bar v}) $
coordinate the horizon locates at ${\bar u} = -4M $.

{{\bf 2. Quantization of Spherically Symmetric Gravity}}

   Let us define the quantum theory of the spherically symmetric gravity. The
partition function is expressed in terms of the path integral over the two
dimensional metric $g_{\alpha\beta}$, the scale field $\df $ and the matter
fields $f $ as
$$
    Z = \int {D_g(g) D_g(\df) D_g(f) \over {\rm Vol(Diff.)} }
           {\rm e}^{iI_{SSG}(g,\df,f)} ~, \qquad
      I_{SSG} = I_{EH} + I_M ~,
     \eqno(4)
$$
where ${\rm Vol(Diff.)} $ is the gauge volume of diffeomorphism.
The functional measures are defined by the following norms
$$
\eqalign{
    &  < \vg, \vg >_g =
           \int d^2 x \dg g^{\alpha \beta} g^{\gamma \delta}
            (\vg_{\alpha \gamma} \vg_{\beta \delta}
                + \vg_{\alpha\beta} \vg_{\gamma\delta}) ~,        \cr
    &  < \vp, \vp>_g =
           \int d^2 x \dg \vp \vp  ~,                           \cr
    &  < \vf_j , \vf_j >_g =
           \int d^2 x \dg \vf_j \vf_j
             \qquad (j=1, \cdots N)  ~.                       \cr
          }   \eqno(5)
$$

    The measures explicitlly depend on the dynamical field $g $. Therefore
we must extract its contributions from the measures. They are evaluated by
using the procedure  of David-Distler-Kawai (DDK)\refmark{\dk} in conformal
gauge $g=\bgr$, where $\bg $ is the background metric. The final expression is
given by
$$
      Z = \int D_{\bg} (\Phi) \e^{i{\hat I}(\bg,\Phi)} ~,
         \eqno(6)
$$
where $\Phi $ denotes the fields $\rho, \df, f$ and the reparametrization
ghosts $ b $ and $c $. The gauge-fixed action ${\hat I} $ is
$$
\eqalign{
  {\hat I} & = \kappa_s  S_L (\rho, \bg) + I_{EH}(\bgr,\df)
              + I_M (\bg,f) + I_{gh} (\bg,b,c)                        \cr
          & = {1 \over 2}  \int d^2 x \dbg \Bigl[
     {\hat g}^{\alpha \beta} \partial_{\alpha} \df \partial_{\beta} \df
    +2{\hat g}^{\alpha \beta} \df \partial_{\alpha} \df \partial_{\beta} \rho
       + {1 \over 2} \bcv \df^2 + \e^{2\rho}          \cr
      & \qquad\qquad + \kappa_s (
         {\hat g}^{\alpha \beta} \partial_{\alpha} \rho \partial_{\beta} \rho
         + \bcv \rho )
       - \sum^N_{j=1}
         {\hat g}^{\alpha \beta} \partial_{\alpha} f_j \partial_{\beta} f_j
       \Bigr] +I_{gh}(\bg,b,c)                             \cr
        }  \eqno(7)
$$
with
$$
      \kappa_s ={1 \over 12\pi}(1+c_{\df}+N-26)={N-27 \over 12\pi} ~,
        \eqno(8)
$$
where $S_L (\rho,\bg) $ is the Liouville action defined by
$$
     S_L (\rho,\bg) = {1 \over 2} \int d^2 x \dbg (
        {\hat g}^{\alpha \beta} \partial_{\alpha} \rho \partial_{\beta} \rho
         + \bcv \rho ) ~.
       \eqno(9)
$$
The value of $\kappa_s $ is given by setting $\xi =1/2 $ in ref.6.
The functional measure of the Liouville field $\rho $ is defined by the norm
on $\bg $ as
$$
     <\vr,\vr>_{\bg} =\int d^2 x \dbg \vr \vr
       \eqno(10)
$$
and also the measures for $\df $ and $f $ is defined by the norms (5) on
$\bg $ instead of $g $.

  The background metric $\bg $ is very artificial so that the theory should
be independent of how to choose it. Really it is proved that the partition
function is invariant under the conformal change of the background metric,
or $Z(\bgs )=Z(\bg ) $, where $\sigma $ is an arbitrary local function.
This means that the theory is considered as a kind of conformal field theory
defined on $\bg $. The Virasoro algebra without central extension should be
realized.\footnote\star{Note that in this case the theory does not reduce to
the free-like theory. So it is quite different from the usual conformal field
theory.}
  The physical state conditions are derived from the demand that the theory
should be independent of how to choose the background metric,
$$
   {\delta Z \over \delta {\hat g}^{\alpha \beta} }
        \biggr\vert_{\bg=\eta} = 0 ~, \quad \hbox{or} \qquad
      <{\hat T}_{\alpha \beta}> = 0 ~,
    \eqno(11)
$$
where $\eta_{\alpha\beta} =(-1,1) $ and the energy-momentum tensor
is defined by
${\hat T}_{\alpha \beta}=-{2 \over \dbg}{\delta {\hat I} \over \delta
\bg^{\alpha\beta}} \vert_{\bg=\eta}$.

  Since the Liouville field $\rho $ is transformed as
$\rho^{\pp}(x^{\pp}) = \rho(x)- \gamma(x) $ for the conformal coordinate
transformation $x^{\pm\pp}=x^{\pm\pp}(x^{\pm}) $,
where $\gamma(x)={1 \over 2}\log \vert {\pd x^{\pp} \over \pd x} \vert^2 $,
$\vert x \vert^2 =x^+ x^- $ and $x^{\pm} =x^0 \pm x^1 $, the energy-momentum
tensor $ {\hat T}_{\alpha \beta} $ is transformed
as\footnote\sharp{More explicitly, ${\hat T}_{\alpha \beta}$ is transformed as
$$
        {\hat T}^{\pp}_{\pm\pm}(x^{\pp})
             = \biggl( {\pd x^{\pm} \over \pd x^{\pm \pp} } \biggr)^2
              \bigl( {\hat T}_{\pm\pm}(x)
                + \kappa_s t_{\pm}(x) \bigr)
                + {c_{tot} \over 12\pi}
                 \biggl( {\pd x^{\pm} \over \pd x^{\pm \pp} } \biggr)^2
                     t_{\pm}(x)
$$
with $c_{tot}=1-12\pi\kappa_s+c_{\df}+N-26=0 $. Note that if
$ {\hat T}^{\pp}_{\pm\pm}(x^{\pp}) $ satisfies the usual form of the Virasoro
algebra with central charge $c_{tot} =0 $, then in $x$ coordinate  the
combination ${\hat T}_{\pm\pm}(x) +
\kappa_s t_{\pm}(x)$, not ${\hat T}_{\pm\pm}(x) $ itself, just satisfies the
same form of the Virasoro algebra.
The importance of $t_{\pm} $ in quantum gravity is stressed in ref.10.}
$$
         {\hat T}^{\pp}_{\pm\pm}(x^{\pp})
             = \biggl( {\pd x^{\pm} \over \pd x^{\pm \pp} } \biggr)^2
              \bigl( {\hat T}_{\pm\pm}(x)
                    + \kappa_s  t_{\pm}(x) \bigr)  ~,
         \qquad {\hat T}^{\pp}_{+-}(x^{\pp})
             = \biggl\vert {\pd x \over \pd x^{\pp}} \biggr\vert^2
              {\hat T}_{+-}(x)   ~,
           \eqno(12)
$$
where $t_{\pm}(x)$ is the Schwarzian derivative
$  t_{\pm}(x) =  (\pd \gamma(x) / \pd x^{\pm} )^2 -
\pd^2 \gamma(x) / \pd x^{\pm 2}$. $t_{\pm} $ is determined by the boundary
condition that the coordinate system which is joined to the Minkowski space
time (asymptotically) is considered as the coordinate system with
$t_{\pm}=0$.

{{\bf 3. Black hole dynamics}}

   To derive the black hole dynamics we must solve the physical state
conditions (11). But it is a very difficult problem
so that we take an approximation. The original actions (1) and (2) are the
 order of $1/\hbar $, while the Liouville part of ${\hat I}$ is the zeroth
order of $\hbar $. However, if $\vert \kappa_s \vert $ is large enough, then
it is meaningful to consider the \lq\lq classical" dynamics of ${\hat I}$.
This is nothing but the semi-classical approximation, which is valid only in
the case of $M \gg 1 $ and $\vert \kappa_s \vert \gg 1$.  The classical
dynamics of ${\hat I}$ is ruled by the equations ${\hat T}_{\alpha\beta} =0 $
and
the $\df $ field equation of motion. Then we set
${\hat T}^{gh}_{\alpha\beta} =0 $ because the ghost flux should vanish in the
flat space time.

    The gravitational collapse geometry is given as a solution with non-zero
infalling matter flux. Giving the flux
${\hat T}^f_{{\bar v}{\bar v}} = M \delta ({\bar v}) $, we can get the exact
solution along the shock wave line ${\bar v}=0 $,
$$
      \pd_{{\bar v}} \df ({\bar v}=0 ,{\bar u}) =
         {1 \over 2} \biggl(
             1 - {4M  \over \sqrt{{\bar u}^2 - 4\kappa_s} } \biggr) ~, \qquad
       \df ({\bar v}=0, {\bar u}) =r=-{1 \over 2}{\bar u} ~.
       \eqno(13)
$$
The (apparent) horizon, which is defined by the equation
$\pd_{{\bar v}} \df =0 $,\refmark{\tih} locates at
$$
     {\bar u} = -4M \sqrt{1+{\kappa_s \over 4M^2 }} ~, \qquad {\bar v}=0 ~.
    \eqno(14)
$$
If $\kappa_s >0 $, the location of the horizon initially shifts to the outside
of the classical horizon ${\bar u}=-4M $ by quantum effects. Then the black
hole evaporates and the horizon approaches to the singularity asymptotically.
The location of the singularity is determined by the equation
$\df^2 =\kappa_s $ (at ${\bar v}=0 $, it is ${\bar u}=-2 \sqrt{\kappa_s}$).
Note that at the singularity the curvature is singular, but the metric is
finite.  If $\kappa_s < 0 $, the singularity disappears. The location  of the
horizon initially shifts to the inside of the classical horizon. If the
effective mass of the black hole  is defined
by $M_{BH} = {1 \over 2} \df \vert_{\hbox{horizon}}$, this means that the
initial mass of the black hole is less than the infalling matter flux $M $.
After the black hole is formed, the positive flux comes in through the horizon
and the black hole mass increases. It seems that the horizon approaches to the
classical horizon asymptotically and becomes stable.

  For $\kappa_s >0 $, the classically forbidden region $\kappa_s > \df^2 >0 $
called \lq\lq Liouville region" extends behind the singularity. To understand
this region we must go back to the full quantum gravity. In the canonical
quantization the physical state conditions are written as
$$
         {\hat T}_{00} \Psi = {\hat T}_{01} \Psi =0 ~,
           \eqno(15)
$$
where $\Psi $ is a physical state and
$$
\eqalign{
   {\hat T}_{00}  = & {1 \over \df^2 -\kappa_s}
            \Bigl( {1 \over 2}\Pi^2_{\rho} -\df \Pi_{\df} \Pi_{\rho}
                  + {\kappa_s \over 2} \Pi^2_{\df} \Bigr)
           +  \df \df^{\prime\prime} + {1 \over 2} \df^{\pp 2}
                 -\df \df^{\prime} \rho^{\prime}
                 -{1 \over 2} \e^{2\rho}                   \cr
           & \qquad\qquad \qquad -{\kappa_s \over 2}
               \bigl( \rho^{\prime 2} -2\rho^{\prime\prime} \bigr)
           + {1 \over 2} \sum^N_{j=1}  \Bigl(  \Pi^2_{f_j}
                   +  f^{\prime 2}_j \Bigr) ~,                     \cr
  {\hat T}_{01} = &
         \rho^{\pp} \Pi_{\rho} - \Pi^{\pp}_{\rho}
            + \df^{\pp} \Pi_{\df} + \sum^N_{j=1} \Pi_{f_j} f^{\pp}_j ~. \cr
       }  \eqno(16)
$$
These correspond to the Hamiltonian and the momentum constraints.
The conjugate momentums for $\rho, \df $ and $f_j $ are defined by
$$
     \Pi_{\rho} = -\kappa_s {\dot \rho} - \df {\dot \df} ~, \quad
     \Pi_{\df} = - {\dot \df} - \df {\dot \rho}  ~,  \quad
     \Pi_{f_j} =  {\dot f}_j  ~.
         \eqno(17)
$$
The notable point is the factor $(\df^2 -\kappa_s )^{-1} $ in front of the
kinetic term of the Hamiltonian constraint, which does not appear in ADM
formalism.\refmark{\tih} The region $\df^2 >\kappa_s $ is classically allowed,
whereas the
Liouville region $\kappa_s >\df^2 > 0 $ is the classically forbidden region
where the sign of the kinetic term changes. There may be some possibility of
gravitational tunnelings through this region.

   The problem of the information loss seems to come out in the case of
$\kappa_s > 0 $.
Then the black hole evaporates and the information seems to be lost.
However in this case the Liouville region extends behind the singularity.
So it appears that  there is a possibility that
the informations run away through this region by gravitational tunneling.
On the other hand, if $\kappa_s \leq 0 $, the Liouville region disappears.
But the black hole seems to be stable. In this case it appears that the
problem of the information loss does not exist.

{{\bf 4. Discussions}}

   The quantum model of spherically symmetric gravity discussed in this talk
has some problems. Here we adopt the conformal matter described by the
action (2). Strictly speaking, however, we should consider the action
such as $I_M = - {1 \over 2} \int d^2 x \sqrt{-g} \df^2 g^{\alpha\beta}
\pd_{\alpha} f \pd_{\beta} f$, which is derived by reducing the four
dimensional action to the two dimensional one. Ignoring $\df^2 $-factor
corresponds to ignoring
the potential which appears when we rewrite the d'Alembertian in terms of the
spherical coordinate. The black hole dynamics is determined by the behavior
near the horizon so that it seems that this simplification does not change
the nature of dynamics.

   The other problem is in the definitions of measures. As the actions
are derived from the four dimensional ones, the two dimensional measures also
should be derived from the four dimensional one
$$
      <\delta g^{(4)},\delta g^{(4)}>_{g^{(4)}}
           = \int d^4 x \hbox{$\sqrt{-g^{(4)}}$} g^{(4)ab}g^{(4)cd}
           (\delta g^{(4)}_{ac}\delta g^{(4)}_{bd}
           +  \delta g^{(4)}_{ab}\delta g^{(4)}_{cd} ) ~.
       \eqno(18)
$$
{}From this definition we get
$$
\eqalign{
    &  < \vg, \vg >_g =
           \int d^2 x \dg \df^2 g^{\alpha \beta} g^{\gamma \delta}
            (\vg_{\alpha \gamma} \vg_{\beta \delta}
                +  \vg_{\alpha\beta} \vg_{\gamma\delta}) ~,        \cr
    &  < \vp, \vp>_g =
           \int d^2 x \dg \vp \vp  ~.                           \cr
        } \eqno(19)
$$
And also for the matter fields,
$$
      < \vf_j , \vf_j >_g =
           \int d^2 x \dg \df^2 \vf_j \vf_j
             \qquad (j=1, \cdots N)  ~.
       \eqno(20)
$$
The difference between (5) and (19-20) is apparent. The factor $\df^2 $ in
the measures of $g $ and $f $ prevents us from quantizing the spherically
symmetric gravity exactly. We expect that this factor also does not change the
nature of quantum dynamics drastically.

\refout

\bye